\documentclass[twocolumn,aps,prb]{revtex4}
\usepackage{amsmath}
\usepackage{amssymb}

\usepackage{graphicx}

\begin{document}

\title{Non-Ohmic critical fluctuation Hall conductivity of layered superconductors
in strong electric fields}

\author{I. Puica}

\email{ipuica@ap.univie.ac.at}

\author{W. Lang}

\affiliation{Institut f\"{u}r Materialphysik der Universit\"{a}t Wien, Boltzmanngasse
5, A-1090 Wien, Austria}

\begin{abstract}
The excess Hall conductivity, resulting from thermal fluctuations
of the superconducting order parameter, is calculated for a layered
superconductor for an arbitrarily strong in-plane electric field and
a perpendicular magnetic field in the frame of the time-dependent
Ginzburg-Landau theory. The fluctuation Hall conductivity is suppressed
in high electric fields much stronger than the longitudinal one. For
high-temperature superconductors we predict a pronounced non-Ohmic
behavior of the excess Hall effect near the critical temperature in
moderate magnetic fields and electric fields of the order of 100 V/cm.
\end{abstract}

\pacs{74.20.De,74.25.Fy,74.40.+k}

\maketitle
The non-Ohmic behavior of the fluctuation conductivity in strong electric
fields, studied for the first time for the isotropic case in connection
with the low temperature superconductors,\cite{Schmid69,Tsuzuki70}
can be summarized by saying that reasonably high values of the electric
field can accelerate the fluctuating paired carriers to the depairing
current, and thus, suppress the lifetime of the fluctuations, which
leads to a deviation from Ohm's law. For a layered superconductor,
a situation very much resembling the crystal structure of the high
temperature superconductors (HTSC), the issue has been addressed theoretically
starting from a microscopic approach\cite{Varlamov92} and subsequently
in the frame of the time-dependent Ginzburg-Landau (TDGL) theory,
in the Gaussian\cite{Mishonov02} as well as in the self-consistent
Hartree approximation.\cite{PuicaLangE} The non-Ohmic fluctuation
conductivity was studied recently also in the presence of a perpendicular
magnetic field,\cite{PuicaLangM} in the Hartree approximation for
the TDGL theory, revealing that the simultaneous application of electric
and magnetic fields results only in a slight additional suppression
of the superconducting fluctuations, compared to the case when the
fields are applied individually.

However, a treatment of a possible non-linearity of the \emph{off-diagonal}
(Hall) components of the fluctuation magnetoconductivity tensor in
a high electric field has been lacking yet. The non-vanishing Hall
current due to fluctuating Cooper pairs, ascribed to a hole-particle
asymmetry\cite{FET} and a complex relaxation time in the TDGL theory,
was first calculated in the Gaussian approximation\cite{FET} and
subsequently improved by incorporating the fluctuation interaction
in the frame of a Hartree approach,\cite{UD} or based on the single
particle-hole renormalization.\cite{NE} All the above models for
the so-called excess Hall conductivity represent however linear response
approximations with respect to the longitudinal electric field and
are therefore valid only for small magnitudes of the latter.

Experimentally, several investigations\cite{Kunchur94,Clinton95,Liebich97,Nakao98}
of the Hall effect at high current densities up to $10^{6}$ Acm$^{-2}$
have been reported, but with the aim of overcoming the vortex pinning
and hence of testing its influence on the Hall conductivity. The intrinsic
non-Ohmic effect was neither envisaged nor fortuitously evidenced,
since it needs, as we shall see later, even higher current densities
in order to be unambiguously discerned.

In this paper we shall treat, in the self-consistent Hartree approach,
the thermal fluctuation Hall conductivity for a layered superconductor
in a perpendicular magnetic field and for an arbitrarily strong in-plane
electric field. To our present knowledge this topic has not been treated
yet, not even for the simpler cases of an isotropic superconductor,
or in the Gaussian approximation. For our purpose, we shall adopt
the Langevin approach to the TDGL equation.\cite{Schmid69,UD} Keeping
the same notations as in Ref. \onlinecite {PuicaLangM}, the gauge-invariant
relaxational TDGL equation governing the critical dynamics of the
superconducting order parameter in the $l$-th superconducting plane
writes:\begin{align}
\Gamma_{0}^{-1}\left(1+i\eta\right)\left(\frac{\partial}{\partial t}+2i\frac{eEx}{\hbar}\right)\psi_{l}+a\psi_{l}\nonumber \\
+b\left|\psi_{l}\right|^{2}\psi_{l}-\frac{\hbar^{2}}{2m}\left[\partial_{x}^{2}+\left(\partial_{y}+\frac{2i\, e}{\hbar}xB\right)^{2}\right]\psi_{l}\nonumber \\
+\frac{\hbar^{2}}{2m_{c}s^{2}}(2\psi_{l}-\psi_{l+1}-\psi_{l-1})=\zeta_{l}\left(\mathbf{x},t\right)\;.\label{EQini}\end{align}
 Here $m$ and $m_{c}$ are effective Cooper pair masses in the $ab$-plane
and along the $c$-axis, respectively, $s$ is the distance between
superconducting planes, and $e$ the elementary electric charge. The
order parameter has the same physical dimension as in the three-dimensional
case and SI units are used. The perpendicular magnetic field $B$
is generated by the vector potential $\mathbf{A}=\left(0,xB,0\right)$,
with $x$ and $y$ the in-plane coordinates, and the magnetization
is neglected. The GL potential $a=a_{0}\varepsilon$ is parameterized
by $a_{0}=\hbar^{2}/2m\xi_{0}^{2}=\hbar^{2}/2m_{c}\xi_{0c}^{2}$ and
$\varepsilon=\ln\left(T/T_{0}\right)$, with $T_{0}$ the mean-field
transition temperature, while $\xi_{0}$ and $\xi_{0c}$ are the in-plane
and out-of-plane coherence lengths extrapolated at $T=0$, respectively.
The real part of the relaxation time in the TDGL equation is given\cite{Masker69}
by $\Gamma_{0}^{-1}=\pi\hbar^{3}/16m\xi_{0}^{2}k_{B}T$, while the
imaginary part $\Gamma_{0}^{-1}\eta$ must be introduced in order
to break the particle-hole symmetry and allow for a non-vanishing
Hall current.\cite{FET,UD,NE} The Langevin white-noise forces $\zeta_{l}\left(\mathbf{x},t\right)$
are correlated through $\left\langle \zeta_{l}\left(\mathbf{x},t\right)\zeta_{l'}^{*}\left(\mathbf{x}',t'\right)\right\rangle =2\Gamma_{0}^{-1}k_{B}T\delta(\mathbf{x}-\mathbf{x}')\delta(t-t')\delta_{ll'}/s$,
where $\delta(\mathbf{x}-\mathbf{x}')$ is the 2-dimensional delta-function
concerning the in-plane coordinates. The electric field $\mathbf{E}$
is assumed along the $x$-axis, generated by the scalar potential
$\varphi=-Ex$. In the chosen gauge, the fluctuation current density
along the $y$ direction in the $l$-th plane, averaged with respect
to the noise, writes: \begin{eqnarray}
\left\langle j_{y}^{(l)}\right\rangle  & = & \left.\frac{ie\hbar}{m}(\partial_{y}-\partial_{y'})\left\langle \psi_{l}\left(x,y,t\right)\psi_{l}^{*}\left(x,y',t\right)\right\rangle \right|_{y=y'}\nonumber \\
 &  & -\frac{4e^{2}}{m}xB\left\langle \left|\psi_{l}\left(x,y,t\right)\right|^{2}\right\rangle \;,\label{CurrentDef}\end{eqnarray}
 so that the fluctuation Hall conductivity is given by $\Delta\sigma_{xy}=-\Delta\sigma_{yx}=-\left\langle j_{y}^{(l)}\right\rangle /E$.

As mentioned, the quartic term in the thermodynamical potential will
be treated in the Hartree approximation,\cite{UD,Penev0} which results
in a linear problem with a modified (renormalized) reduced temperature
$\widetilde{\varepsilon}=\varepsilon+b\left\langle \left|\psi_{l}\right|^{2}\right\rangle /a_{0}$.

Following the same procedure as in Ref. \onlinecite
{PuicaLangM}, we introduce the Fourier transform with respect to the in-plane
coordinate $y$, the layer index $l$, and time $t$, respectively,
and also the Landau level (LL) representation with respect to the
$x$-dependence, through the relation:\begin{align}
\psi_{l}(x,y,t) & =\int\frac{dk}{2\pi}\int_{-\pi/s}^{\pi/s}\frac{dq}{2\pi}\int\frac{d\omega}{2\pi}\sum_{n\geq0}\psi_{q}(n,k,\omega)\label{Fourier}\\
 & \cdot e^{-iky}e^{-iqls}e^{-i\omega t}u_{n}\left(x-\frac{\hbar k}{2eB}\right)\,,\nonumber \end{align}
 where the functions $u_{n}\left(x\right)$ with $n\in\mathbb{N}$
build the orthonormal eigenfunction system of the harmonic oscillator
hamiltonian, so that $\left(-\hbar^{2}\partial_{x}^{2}+4e^{2}B^{2}x^{2}\right)u_{n}\left(x\right)=2\hbar eB\left(2n+1\right)u_{n}\left(x\right)$.

Equation (\ref{EQini}) can be evaluated into the matrix form, after
applying the expansion (\ref{Fourier}): \begin{equation}
a_{0}\sum_{n'}\left(\mathbf{M}+\eta\mathbf{P}\right)_{nn'}\psi_{q}(n',k,\omega)=\zeta_{q}(n,k,\omega)\label{MatrixEq}\end{equation}
 where the new noise terms $\zeta_{q}\left(n,k,\omega\right)$, corresponding
to the expansion rule (\ref{Fourier}), are delta-correlated such
as $\left\langle \zeta_{q_{1}}\left(n_{1},k_{1},\omega_{1}\right)\zeta_{q_{2}}^{*}\left(n_{2},k_{2},\omega_{2}\right)\right\rangle =2\Gamma_{0}^{-1}k_{B}T(2\pi)^{3}\delta(k_{1}-k_{2})\delta(q_{1}-q_{2})\delta(\omega_{1}-\omega_{2})\delta_{n_{1}n_{2}}$,
and where the two dimensionless symmetrical tridiagonal matrices $\mathbf{M}$
and $\mathbf{P}$ have the elements: \begin{eqnarray}
\mathbf{M}_{nn} & = & -i\omega'+\widetilde{\varepsilon}_{nq'}\,;\qquad\mathbf{P}_{nn}=\omega'\,;\nonumber \\
\mathbf{M}_{n+1,n} & = & \mathbf{M}_{n,n+1}=i\, f\sqrt{n+1}\,;\label{MatrixElem}\\
\mathbf{P}_{n+1,n} & = & \mathbf{P}_{n,n+1}=-f\sqrt{n+1}\,;\nonumber \\
\widetilde{\varepsilon}_{nq'} & = & \widetilde{\varepsilon}+\frac{r}{2}\left(1-\cos q'\right)+\left(2n+1\right)h\,;\nonumber \\
r & = & \frac{2\hbar^{2}}{a_{0}m_{c}s^{2}}=\left(\frac{2\xi_{0c}}{s}\right)^{2}\:;\qquad f=2\sqrt{6}\frac{E'}{\sqrt{h}}\,.\nonumber \end{eqnarray}
 Here we have introduced the new variables: \begin{equation}
\omega'=\frac{\Gamma_{0}^{-1}}{a_{0}}\left(\omega-\frac{Ek}{B}\right);\quad q'=qs;\quad k'=\frac{\hbar}{2eB}k\,,\label{NewVariables}\end{equation}
 and the reduced field magnitudes: \begin{equation}
h=\frac{B}{B_{c2}(0)}=\frac{\hbar eB}{ma_{0}}\quad\text{and}\quad E'=\frac{eE\xi_{0}\Gamma_{0}^{-1}}{2\sqrt{3}a_{0}\hbar}=\frac{E}{E_{0}}\;,\label{ReducedFields}\end{equation}
 with $E_{0}=16\sqrt{3}k_{B}T\,/\,\pi e\xi_{0}$ defined as in Refs.
\onlinecite{Varlamov92} and \onlinecite {Mishonov02}.

By solving Eq. (\ref{MatrixEq}), and taking into account the expansion
form (\ref{Fourier}), one obtains the correlation function of the
order parameter: \begin{eqnarray}
 &  & \left\langle \psi_{l}\left(x,y,t\right)\psi_{l}^{*}\left(x,y',t\right)\right\rangle =\frac{4mk_{B}T}{\hbar^{2}s}h\label{PsiCorr1}\\
 &  & \cdot\int\frac{dk'}{2\pi}\int\frac{d\omega'}{2\pi}\int\frac{dq'}{2\pi}\sum_{n}\sum_{n'}e^{-ik'\frac{2eB}{\hbar}(y-y')}u_{n}\left(x-k'\right)\nonumber \\
 &  & \cdot u_{n'}\left(x-k'\right)\left[\left(\mathbf{M}+\eta\mathbf{P}\right)^{+}\cdot\left(\mathbf{M}+\eta\mathbf{P}\right)\right]_{nn'}^{-1}\left(q',\omega'\right)\,,\nonumber \end{eqnarray}
 where the notation $\left[...\right]_{nn'}^{-1}$ is to be understood
as the element of the inverted matrix.

Before proceeding further we point out that the sums over the LL in
Eqs. (\ref{Fourier}), (\ref{MatrixEq}) and (\ref{PsiCorr1}) must
be cut-off at some index $N_{c}$, reflecting the inherent UV divergence
of the Ginzburg-Landau theory. The classical\cite{Schmid69,Penev0}
procedure is to suppress the short wavelength fluctuating modes through
a \emph{momentum} (or, equivalently, \emph{kinetic energy}) \emph{cut-off}
condition, which, in terms of the LL representation writes\cite{UD,Penev0}
$\left(\hbar e_{0}B/m\right)\left(n+\frac{1}{2}\right)\leq ca_{0}=c\hbar^{2}/2m\xi_{0}^{2}$,
with the cut-off parameter $c$ of the order of unity. A \emph{total
energy cut-off} was also recently proposed,\cite{Vidal02} whose physical
meaning was shown to follow from the uncertainty principle. However,
in the critical fluctuation region the two cut-off conditions almost
coincide quantitatively, due to the low reduced-temperature $\varepsilon$
with respect to $c$, so that we shall apply for simplicity the momentum
cut-off procedure. In terms of the reduced magnetic field $h$, it
writes thus $h\left(N_{c}+\frac{1}{2}\right)=c/2$. In this way, the
matrices $\mathbf{M}$ and $\mathbf{P}$ are truncated at $N_{c}+1$
lines and columns.

The value of $\eta$ can be inferred from the microscopical theory
if one considers the energy derivative $\mathcal{N}'$ of the density
of states $\mathcal{N}$ at the Fermi level $\varepsilon_{F}$, and
it writes\cite{NE} $\eta=-\left(k_{B}T/\varepsilon_{F}\right)\alpha$,
where the parameter $\alpha$ amounts in the BCS model to $\alpha=4\varepsilon_{F}\mathcal{N}'/\pi g_{\mathrm{BCS}}\mathcal{N}^{2}$,
with $g_{\mathrm{BCS}}$ ($>0$) the BCS coupling constant.\cite{FET,NE}
Since $\varepsilon_{F}$ is for HTSC of the order of $10^{3}$ K (in
$k_{B}$ units),\cite{Rietveld92} and the hole-particle asymmetry
parameter $\alpha$, inferred from fits of excess Hall effect data\cite{Lang94,SDiego03}
with the models from Refs. \onlinecite{FET} and \onlinecite{NE},
turns out to be of the order of $10^{-2}\div10^{-1}$, we conclude
that $\eta$ is a small parameter, reflecting also the small Hall
angle. We shall therefore expand the inverted matrix in Eq. (\ref{PsiCorr1})
only up to the linear term in $\eta$, such as $\left[\left(\mathbf{M}+\eta\mathbf{P}\right)^{+}\cdot\left(\mathbf{M}+\eta\mathbf{P}\right)\right]^{-1}=\mathbf{Q}-\eta\mathbf{Q}\cdot\mathbf{K}\cdot\mathbf{Q}+\mathcal{O}\left(\eta^{2}\right)$,
where the Hermitian matrix $\mathbf{Q}=\left(\mathbf{M}^{+}\cdot\mathbf{M}\right)^{-1}$,
and the symmetrical tridiagonal matrix $\mathbf{K}$ has the elements:
$\mathbf{K}_{nn}=2\omega'\,\widetilde{\varepsilon}_{nq'}\,;\;\mathbf{K}_{n+1,n}=\mathbf{K}_{n,n+1}=-f\sqrt{n+1}\left(\widetilde{\varepsilon}_{nq'}+\widetilde{\varepsilon}_{n+1,q'}\right)$.

By using the correlation function (\ref{PsiCorr1}) in the current
density expression (\ref{CurrentDef}), we can eventually write the
fluctuation Hall conductivity in the form: \begin{widetext}\begin{eqnarray}
\Delta\sigma_{xy} & = & \eta\frac{e^{2}}{\hbar s}\,\frac{h}{f}\,\int_{-\infty}^{\infty}\frac{d\omega'}{2\pi}\int_{-\pi}^{\pi}\frac{dq'}{2\pi}\sum_{n=0}^{N_{c}-1}\sqrt{n+1}\:\mathrm{Re}\left[-2\omega'\sum_{n'=0}^{N_{c}}\mathbf{Q}_{n+1,n'}\,\widetilde{\varepsilon}_{n'q'}\mathbf{Q}_{n',n}\right.\nonumber \\
 &  & \left.+2f\sum_{n'=0}^{N_{c}-1}\sqrt{n'+1}\,\widetilde{\varepsilon}_{n'+\frac{1}{2},q'}\left(\mathbf{Q}_{n+1,n'}\mathbf{Q}_{n'+1,n}+\mathbf{Q}_{n+1,n'+1}\mathbf{Q}_{n',n}\right)\right]\label{SigmaH}\end{eqnarray}
\end{widetext}where we have explicitly specified the cut-off in the
Landau level sum. We note also that the zeroth order in $\eta$ gives
no contribution to $\Delta\sigma_{xy}$, as also established on the
general grounds of the Onsager relations.\cite{UD,NE} The electric
field enters Eq. (\ref{SigmaH}) through the parameter $f$, defined
in Eqs. (\ref{MatrixElem}). In order to apply the expression for
$\Delta\sigma_{xy}$ also in the limit $E\rightarrow0$, one has to
expand the $\mathbf{Q}$-matrix elements up to the linear term in
$f$, namely $\mathbf{Q}=\mathbf{Q}^{(0)}+f\,\mathbf{Q}^{(1)}+\mathcal{O}\left(f^{2}\right)$.
Since $\mathbf{Q}^{(0)}$ is diagonal and has the elements $\mathbf{Q}_{nn}^{(0)}=\left\{ \omega'^{2}+\widetilde{\varepsilon}_{nq'}^{2}\right\} ^{-1}$,
one needs from $\mathbf{Q}^{(1)}$ only the elements $\mathbf{Q}_{n+1,n}^{(1)}=2\sqrt{n+1}\mathbf{Q}_{nn}^{(0)}\mathbf{Q}_{n+1,n+1}^{(0)}\left(\omega'-ih\right)$
and finally obtains in the linear response approximation:\begin{equation}
\Delta\sigma_{xy}^{(0)}=\frac{\eta e^{2}h^{3}}{2\hbar s}\int_{-\pi}^{\pi}\frac{dq'}{2\pi}\sum_{n=0}^{N_{c}-1}\frac{n+1}{\widetilde{\varepsilon}_{nq'}\widetilde{\varepsilon}_{n+1,q'}\widetilde{\varepsilon}_{n+\frac{1}{2},q'}^{2}}\label{SigmaH0}\end{equation}
 which matches the formula found in Ref. \onlinecite{UD} and is also
coincident, after performing the $q'$-integral and replacing $\eta=-\left(k_{B}T/\varepsilon_{F}\right)\alpha$,
with the expression given by Ref. \onlinecite{NE}. If one neglects
the cut-off procedure (i.e. makes $N_{c}\rightarrow\infty$), and
also the renormalization (i.e. replaces $\widetilde{\varepsilon}$
by $\varepsilon$), the result (\ref{SigmaH0}) agrees with the Hall
conductivity expression in weak electric field and arbitrary magnetic
field found from microscopical calculations in Ref. \onlinecite{Larkin02}.\footnote{In Ref. \onlinecite{Larkin02} a factor $2/\pi$ seems to be missing in the general formula for Hall conductivity, and a factor $1/\pi$ in the limit expression for weak magnetic field.}
Taking further also the limit of weak magnetic field, one can find
the expression given by Ref. \onlinecite{Mishonov03} in the framework
of the Boltzmann kinetic equation.

Starting from Eq. (\ref{PsiCorr1}) we are able to compute also the
fluctuation Cooper pair density $\left\langle \left|\psi\right|^{2}\right\rangle $,
keeping only the dominant zero-th order term in $\eta$, and write
the self-consistent equation for the renormalized reduced temperature
parameter $\widetilde{\varepsilon}$, as already found in Ref. \onlinecite
{PuicaLangM}: \begin{equation}
\widetilde{\varepsilon}=\ln\frac{T}{T_{0}}+gT\,4h\int_{-\infty}^{\infty}\frac{d\omega'}{2\pi}\int_{-\pi}^{\pi}\frac{dq'}{2\pi}\sum_{n}\mathbf{Q}_{nn}\left(q',\omega'\right)\,.\label{self-const-eq}\end{equation}
 Here, we have introduced the parameter $g=2\mu_{0}\kappa^{2}e^{2}\xi_{0}^{2}k_{B}/\left(\pi\hbar^{2}s\right)$
and we have taken into account the expression of the quartic term
coefficient\cite{UD} $b=2\mu_{0}\kappa^{2}e^{2}\hbar^{2}/m^{2}$,
with $\kappa$ the Ginzburg-Landau parameter $\kappa=\lambda_{0}/\xi_{0}$.
In analogy with the Gaussian fluctuation case, we shall adopt as definition
for the critical temperature $T_{c}(E,B)$ the vanishing of the reduced
temperature, $\widetilde{\varepsilon}=0$. The relationship between
$T_{0}$ and $T_{c}(0,0)\equiv T_{c0}$, corresponding to Eq. (\ref{self-const-eq})
taken in the zero-fields limit at $\widetilde{\varepsilon}=0$, has
been already found in Ref. \onlinecite{PuicaLangE} and writes $T_{0}=T_{c0}\left[\sqrt{c/r}+\sqrt{1+(c/r)}\right]^{2gT_{c0}}$.

We shall take as example the optimally doped YBa$_{2}$Cu$_{3}$O$_{7-x}$,
for which typical characteristic parameters are: $s=1.17$ nm, $\xi_{0}=1.2$
nm, $\xi_{0c}=0.14$ nm, $\kappa=70$ and $T_{c0}=92$ K. We assume
a positive (hole-like) normal-state Hall conductivity $\sigma_{xy}^{N}$
obeying the Anderson's formula $\sigma_{xx}^{N}/\sigma_{xy}^{N}=A\, T^{2}$
with a generic $A=0.07$ K$^{-2}$ at $B=1$ T, and a linear extrapolation
for the normal state resistivity vanishing at $T=0$, with a typical
value $\rho_{xx}^{N}=84\,\mu\Omega$cm at $T=200$ K. The cut-off
parameter $c=1$. For the Fermi energy we take $\varepsilon_{F}=k_{B}\cdot10^{3}$
K,\cite{Rietveld92} while the parameter $\alpha$ will be given the
positive value $\alpha=0.01$,\cite{Lang94,Hall04} in order to have
$\eta<0$ and $\Delta\sigma_{xy}<0$, and thus to account for the
Hall effect's sign change occuring in the transition region.

It should be mentioned that the relevance of the $\mathcal{N}'$-sign
to the sign change of the Hall effect is still open to debate. The
conventional $s$-wave weak coupling BCS theory predicts a positive
excess Hall effect in the underdoped cuprates for a hole-like Fermi
surface, in contrast to the experimental reports,\cite{Nagaoka98}
although recent theoretical approaches based on the presence of preformed
pairs,\cite{Geshkenbein97} on an additional contribution to the particle-hole
asymmetry coming from the quadratic electron spectrum\cite{Sergeev02}
or on the proximity of an electronic topological transition\cite{Angilella03}
point out the possibility of an electron-like Hall sign in the hole-like
doping range. Our purpose is however to illustrate the high electric
field effect on the fluctuation Hall conductivity, for which $\alpha$
(or, equivalently, $\eta$) represents merely a prefactor that could
be inferred from fits to the measurements. The experimental study
of the non-Ohmic Hall effect could provide thus a supplementary and
better tool for assessing the particle-hole asymmetry parameter $\alpha$,
without the uncertainty introduced by the previous\cite{Lang94,Hall04}
need to estimate the background normal state contribution $\sigma_{xy}^{N}$,
because the difference $\sigma_{xy}\left(E\right)-\sigma_{xy}\left(0\right)=\Delta\sigma_{xy}\left(E\right)-\Delta\sigma_{xy}\left(0\right)$
would be independent of $\sigma_{xy}^{N}$. An eventual experimental
evidence of the non-Ohmic Hall effect behavior would also bring a
strong argument in favour of the superconducting fluctuations in the
long-lasting debate on the causes of the Hall effect sign change in
HTSC.

\begin{figure*}
\includegraphics[  width=15cm]{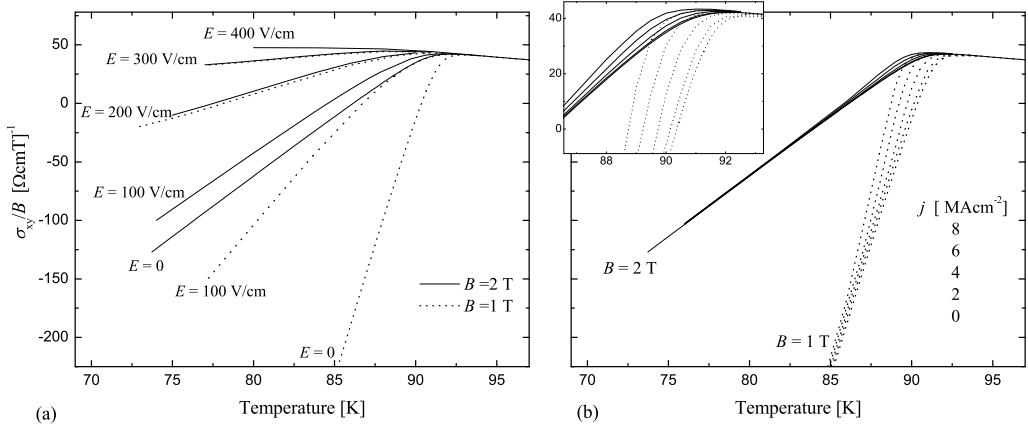}

\caption{Hall conductivity normalized to the magnetic field $B$ as a function
of temperature for two values of the magnetic field, at several magnitudes
of the (a) in-plane electric field; or (b) current density. A detail
is shown in the inset.\label{Fig1}}
\end{figure*}

Figure \ref{Fig1}a shows the Hall conductivity $\sigma_{xy}=\sigma_{xy}^{N}+\Delta\sigma_{xy}$,
normalized to the magnetic field value, for $B=1$ T (dotted lines)
and $B=2$ T (solid lines) at different magnitudes of the in-plane
electric field. One can notice a strong suppression of the fluctuation
contribution to the Hall conductivity $\Delta\sigma_{xy}$ due to
the high electric field. A higher magnetic field leads to a stronger
reduction of the fluctuation Hall conductivity \emph{per se} and thus,
the fluctuation suppression as a function of the electric field becomes
relatively smaller. For $E>400$ V/cm the fluctuation part $\Delta\sigma_{xy}$
becomes negligible with respect to the normal-state component $\sigma_{xy}^{N}$,
so that the Hall conductivity turns out to be much more sensitive
to high electric fields than the longitudinal one $\sigma_{xx}$.
The latter preserves a significant superconducting fluctuation component
even for fields $E=1500$ V/cm and $B=11$ T, as shown in Ref. \onlinecite{PuicaLangM}.
This might be not too surprising, since $\sigma_{xy}$ is altered
much more than $\sigma_{xx}$ also by a magnetic field alone (Ref.
\onlinecite{Hall04}).

Throughout an experiment, constant current density $j$ can be achieved
much easier than constant $E$. Thus, Fig. \ref{Fig1}b presents the
same non-Ohmic effect on $\sigma_{xy}$ with $j$ as the parameter.
For this purpose, the equation $j=\left(\sigma_{xx}^{N}\left(T\right)+\Delta\sigma_{xx}\left(T,E,B\right)\right)E$
was firstly solved with respect to $E$ at fixed $T$, $B$ and $j$,
where for the longitudinal fluctuation conductivity $\Delta\sigma_{xx}$
the result of Ref. \onlinecite{PuicaLangM} has been used:\begin{equation}
\Delta\sigma_{xx}=\frac{e^{2}}{\hbar s}\frac{h}{f}\int_{-\infty}^{\infty}\frac{d\omega'}{2\pi}\int_{-\pi}^{\pi}\frac{dq'}{2\pi}\sum_{n=0}^{N_{c}-1}\sqrt{n+1}\:\mathrm{Im}\left(\mathbf{Q}_{n,n+1}\right)\:.\label{SigmaXX}\end{equation}
 One can see in Fig. \ref{Fig1}b that, due to the very low resistivity
at lower temperatures, high electric field values are difficult to
attain, so that the non-Ohmic effect can be discerned only at the
beginning of the transition, and only for current densities higher
than 2 MAcm$^{-2}$. This explains why the non-Ohmic effect on the
Hall conductivity has not been detected experimentally so far. Nevertheless,
attaining current densities of a few MAcm$^{-2}$, with minimal self-heating,
on cuprate thin films of a typical $d=100$ nm thickness, might not
be such a difficult task, if one uses very short current pulses of
the order of tens of nanoseconds, so that only the phonon mismatch
at the film-substrate interface practically contributes to the sample
temperature rise.\cite{Kunchur95} According to literature data,\cite{Doettinger94,Sergeev94}
a thermal boundary resistance of about $R_{bd}=0.5$ mK$\cdot$cm$^{2}$/W
between YBa$_{2}$Cu$_{3}$O$_{7-x}$ and the substrate MgO would
imply at about 100 K (where $\rho_{xx}^{N}\simeq40\,\mu\Omega$cm)
a temperature rise $\Delta T=dR_{bd}j^{2}\rho_{xx}^{N}\simeq0.8$
K for a current density $j=2$ MAcm$^{-2}$. This accuracy could be
sufficient to discern the non-Ohmic behavior.

In summary, we have calculated the critical fluctuation Hall conductivity
for a layered superconductor in an arbitrary in-plane electric field
and perpendicular magnetic field in the frame of the TDGL theory using
the self-consistent Hartree approximation. The main result is the
formula (\ref{SigmaH}) that was found to reduce to previous results
in the linear response limit Eq. (\ref{SigmaH0}). Qualitatively,
high electric fields result in a strong suppression of the fluctuation
contribution to the Hall conductivity, in particular in moderate magnetic
fields where order-parameter fluctuations are still strong.

This work was supported by the Fonds zur F\"{o}rderung der wissenschaftlichen
Forschung, Austria.

\bibliographystyle{APSREV}
\bibliography{PuicaLang3}

\end{document}